**Ferrovolcanism on metal worlds and the origin of pallasites**

**Authors:** Brandon C. Johnson[1], Michael M. Sori[2], Alexander J. Evans[3]

[1]Department of Earth, Atmospheric, and Planetary Sciences, Purdue University, West Lafayette, IN, USA
[2]Lunar and Planetary Laboratory, University of Arizona, Tucson, AZ, USA

[3]Department of Earth, Environmental and Planetary Sciences, Brown University, Providence, RI, USA

**As differentiated planetesimals cool, their cores can solidify from the outside-in[1], as evidenced by paleomagnetic measurements and cooling rate estimates of iron meteorites[2,3]. The details of outside-in solidification and fate of residual core melt are poorly understood. For a core primarily composed of Fe and Ni alloyed with lighter constituent elements, like sulfur, such inward core growth would likely be achieved by growth of solid FeNi dendrites[4]. Growth of FeNi dendrites results in interconnected pockets of residual melt that become progressively enriched in sulfur up to a eutectic composition of 31 wt% sulfur as FeNi continues to solidify[4]. Here we show that regions of residual sulfur-enriched FeNi melt in the core attain sufficient excess pressures to propagate via dikes into the mantle. Thus, core material will intrude into the overlying rocky mantle or possibly even erupt onto the plantesimal's surface. We refer to these processes collectively as "ferrovolcanism". Our calculation show that ferrovolcanic surface eruptions are more likely on bodies with mantles less than 50 km thick. We show that intrusive ferromagmatism can produce pallasites, an enigmatic class of meteorites composed of olivine crystals entrained in a matrix of FeNi metal[4]. Ferrovolcanic eruptions may explain the observations that Psyche has a bulk density inconsistent with iron metorites[5] yet shows evidence of a metallic surface composition[6].**

We calculate the excess pressure of residual core melt to determine when intrusive ferromagmatism will occur. When a melt is less dense than the surrounding solid material, an interconnected region of melt of vertical extent, *h* (Figure 1), produces an excess pressure $P_e = \Delta\rho h g$, where $\Delta\rho$ is the density contrast between solid and melt, and *g* is the gravitational acceleration (see Methods)[7]. The density contrast depends on the sulfur content of the melt and the gravitational acceleration is set by the size of the core (Figure 2). Simulations of inward core solidification by ref. [4] on a 200-km-diameter mantle-stripped core produce a concentric partially-molten layer composed of solid FeNi dendrites and sulfur-enriched liquid (melt) that is approximately 10–30 km thick and the largest interconnected regions of melt are likely similar in scale. Self-propagating fluid filled cracks called dikes form when excess pressures exceed the tensile strength of rock ~0.5–6 MPa (ref.[8], Methods). Our calculations show that sufficient excess pressures can be achieved under a range of reasonable conditions (Figure 2). For example, for vertical extents of melt of 5 km, sulfur content of 30 wt%, and a core radius of 50 km, the excess pressures is 2 MPa. Higher excess pressures occur for larger cores and larger vertical extents of melt, while lower excess pressures occur at lower sulfur contents (Figure 2). If the core is too large, however, it is unlikely that the core would grow from the outside-in, and these calculations would no longer apply[1,4].

If the top of the partial melt region lies directly underneath the rocky mantle or if surrounding iron is sufficiently warm, with an expected temperature ranging from 1261 to 1811 K (Methods), excess pressures of a few MPa should be capable of causing dikes to propagate into the mantle of the planetesimal (see Methods). It is important to note that ascent and eruption can occur even when the magma is much denser than the surrounding material. In this case ascent is driven by excess pressure rather than buoyancy of the magma (e.g. lunar mare basalt magmas, which were on average 350–460 kg/m$^3$ more dense than the surrounding 20 km thick lunar crust they ascended through)[7,9,10]. Our calculations do not consider stress concentrations that are expected for the angular shape of melt pockets produced by dendritic growth or the pre-existing tectonic stress states, which may lead to conditions that are more conducive to dike propagation[8].

Next, we consider how far such a ferromagmatic dike could propagate into the mantle of a body and whether volcanism (i.e., surface eruptions) might occur. Figure 3 shows an estimate of how far into the overlying mantle of a body a ferromagmatic dike might propagate, assuming the excess pressure supports the negatively buoyant magma (see Methods). Support by excess pressure is in contrast to eruption of Fe, Ni-FeS eutectic melts driven by buoyancy provided by entrained bubbles considered by ref.[11]. If this penetration height exceeds the mantle thickness, surface eruptions are expected. Although simple, these calculations are relatively conservative and assume the top of the interconnected melt is at the core-mantle interface. Considering pockets of interconnected melt located deeper in the core (Methods), other causes of excess pressure, or pre-existing stress gradients could all lead to conditions more conducive to eruption[9]. Even neglecting these effects, our calculations suggest that ferrovolcanic eruptions may be possible for small, metal-rich bodies, especially for sulfur-rich melts and bodies with mantles thinner than about 35 km (assuming $h$=10), or bodies where the mantle has been locally thinned by large impact craters.

Pallasites are a class of meteorites composed of olivine crystals entrained in a matrix of

FeNi metal[12]. Although pallasite origins are debated, there are several distinct pallasite parent bodies indicating that pallasite formation took place repeatedly in both the inner and outer Solar System[13,14]. Many pallasites have a low iridium content, implying their metal matrix is sourced from a highly evolved (up to ~80%) crystalized metallic melt[15]. Ferromagmatism offers a mechanism to explain pallasites as intrusions of evolved core magmas into the olivine-rich mantle of a body. After intrusion as near vertical dikes and possible lateral spreading as sills[8], the metallic melt would percolate through the olivine matrix before solidifying and producing textures consistent with observed pallasites[12,16]. On average, pallasites are 50% metal by weight[17]. Thus, with an average sulfur content of 2.3 wt%, the metal in main group pallasites may have had a sulfur content of about 5 wt% (ref.[12,17]). The sulfur content in the ferromagma is set by the composition when excess pressure is large enough to cause dike propagation. If residual melts are efficiently removed from the solidifying core when their sulfur content is low, we would not expect to find many pallasites with high sulfur contents. Even at ~5 wt% sulfur, core material could have been intruded several km into the mantle of the pallasite parent body (Figure 3) and possibly farther if the interconnected melt is further pressurized, has a larger vertical extent than assumed here, or pre-existing stress gradients favor ascent[9]. Although the sulfur content of an evolved melt depends on sulfur content of the initial primitive melt, a sulfur content of 5 wt%, is well below the 31 wt% in a eutectic melt. The relatively low sulfur content of pallasites, compared to a eutectic melt, has led some to consider the possibility that sulfur-rich pallasites, like Phillips County[16], are simply underrepresented in the meteorite record[18]. It is possible that the initial ferromagma has greater than 5 wt% sulfur if intrusion of core material occurs before a eutectic composition of 31 wt% sulfur has been reached. In this case, the sulfur content could be progressively enhanced in the residual melt as the intrusion crystallizes. This sulfur-enriched melt may be mobile and leave, producing a region composed of a relatively pure FeNi solid[18] and mantle material.

Paleomagnetic studies show that pallasites record the magnetic field of their parent body[19]. To record such a field, pallasite source material must have been cooled below its Curie temperature (~360 K) while a liquid core was still convecting[19]. Thermal models[19] that include conventional outward core growth suggest that pallasites originated from less than 40 km deep in the mantle (60 km above the core) of a 200-km-diameter body to achieve these low temperatures. Thermal models considering inward core solidification, however, suggest even outer core material may reach temperatures below 360 K while the inner portions of the core remain liquid[4]. Thus, in contrast to constraints from traditional core solidification[19], if core solidification progressed from the outside-in it may be possible that pallasites formed near the core-mantle boundary. Inward core growth also likely affects the lifetime of a core dynamo[4]. Thus, further thermochemical evolution models considering inward core solidification are needed to determine if a ferromagmatic origin for pallasites is consistent with paleomagnetic measurements and thermal constraints[2,19].

Our findings may have also implications for Psyche, an asteroid that is the target of an upcoming NASA spacecraft mission[20]. Psyche was argued to be an intact iron planetary core on the basis of high density and radar albedo measurements[6]. However, the most recent and precise estimate of bulk density 4160±640 kg/m$^3$ (ref.[5]) is much lower than the density of iron meteorites (7000–8000 kg/m$^3$), making this interpretation more uncertain[5]. It has been suggested that Psyche represents the core of an ancient, differentiated planetesimal that was exposed by hit-and-run collisions[21], and contains high macroporosity[5] (47%). A second suggested structure is that Psyche is a differentiated body with a density similar to stony-iron meteorites[5,6].

Ferrovolcanism offers another possible structure to reconcile density measurements and observations of both metal[6] and orthopyroxene[22] on the surface of Psyche. We considered a two-layer structure where a metal core is surrounded by a silicate mantle. Assuming a triaxial ellipsoidal shape (see Methods), a core density of 7500 kg/m$^3$, and a mantle density ranging between 2500–3500 kg/m$^3$, we calculate that an average mantle thickness as low as 25 km is consistent with bulk density estimates (Figure 4). This silicate thickness is compatible with ferrovolcanic eruptions if sulfur content and vertical extent of the partial melt region are sufficiently high (Figure 3). Ferrovolcanism may have transported core material to the surface, causing the radar detections of metal. With low viscosities (~6 mPa s (ref.[23])) and limited dissolved gases[11], we expect that ferrovolcanic eruptions would be effusive fissure eruptions. Unless there is a substantial component of entrained solid rocky material, it is unlikely that a topographic ferrovolcanic edifice would be built during an eruption. Ferrovolcanism would only be expected early in Psyche's geologic history while the core was solidifying. Thus, ferrovolcanic units are likely mixed with rocky material from billions of years of impact bombardment, producing a mixed metal-rock regolith consistent with recent radar observations of a metal-rich surface composition[6]. Observations of gravity and composition by the upcoming Psyche mission should be able to determine if Psyche has a two-layer structure like that in Figure 4. We also note that it is possible the lower mantle of Psyche is dominated by olivine and the upper mantle is dominated by pyroxene, similar to the structure that has recently been suggested for the Moon and Vesta[24]. If spatially resolved spectral observations reveal olivine on the surface of Psyche, this interior structure would also be consistent with the exciting possibility that Psyche is a pallasite parent body.

**Authors Information**

Correspondence and requests for materials should be addressed to B.C.J. (bcjohnson@purdue.edu).

**Acknowledgements:** We thank H. J. Melosh, F. Nimmo, J. N. H. Abrahamson, E. R. D. Scott and M. Caffee for helpful discussion and comments on this work.

**Author Contributions:** B.C.J. conceived this study and application of ferrovolcanism to pallasites and the Psyche observations, which M.M.S. had noted. B.C.J. produced excess pressure and mantle penetration calculation with input from all authors. M.M.S. produced the Psyche density calculations. All authors contributed to preparation of the manuscript and the conclusions presented in this work.

The authors declare no competing financial interests.

**Methods:**

To calculate excess pressure we follow ref.[4] and assume a simplified core composition with solid material as pure iron and the melt as pure Fe-FeS, where the sulfur content of the melt depends on the degree of solidification and original sulfur content of the core. With this assumed composition, excess pressure is $P_e = (\rho_{Fe} - \rho_{FeS})hg$ (ref.7), where $\rho_{Fe} = 7500$ kg/m³ is the assumed density of solid iron4. The density of sulfur rich melt is given by $\rho_{FeS} = 6950 - 5176\chi - 3108\chi^2$ where $\chi$ is atom fraction sulfur, not weight percent25. The gravitational acceleration, $g$, at a given radius $R$ is given by $g = GM/R^2$, where $G$ is the gravitational constant, and $M = \frac{4}{3}\pi R^3 \rho_c$ is the mass of the spherical core. Here we use the density $\rho_c = 7020$ kg/m³ to represent the average density of the liquid core4. The core size in figure 1, $R$, assumes the top of the partial melt region is located at the core mantle boundary. When the sulfur-enriched melt region occurs at greater depth, the core radius in Figure 1 and associated calculations would instead be the radial distance from center of the core, which cannot exceed the core radius.

We expect ferromagmas would propagate through the mantle as dikes. The temperature of the core-mantle boundary will be about equal to the eutectic temperature of Fe-FeS, 1261 K. At this temperature and a stress of a few MPa, the Maxwell time of dry olivine is 180 kyr refs.26,27. Even at the iron liquidus of 1811 K, the Maxwell time is over a month. Thus, we expect failure would be brittle and ferromagma would propagate in dikes. Terrestrial magmas typically have temperatures of 1000–1600 K depending on composition26. Assuming there is minimal superheating, the hottest temperature that we expect for a ferromagma would be is 1811 K, the melt temperature of pure iron, and we expect more evolved melts to have temperatures much closer to the eutectic temperature of 1261 K. These estimates mean that the ferromagma and surrounding rocks would have similar temperatures to the magma and surrounding rocks involved in terrestrial dikes. Terrestrial dikes propagate when excess pressures exceed the tensile strength of rock ~0.5–6 MPa (ref.8) and we expect a similar excess pressures are needed for ferromagmatic dikes to propagate.

In addition to dike propagation in the mantle, we also consider the possibility of dikes propagating through solid iron. The tensile strength of solid iron is 16.8 MPa at 1073 K and is expected to decrease to zero as the melt temperature of 1811 K is approached[28,29]. Below the eutectic temperature of 1261 K, even the sulfur-enriched material is solid. Thus we expect the relevant tensile strength of warm iron above 1261 K will be similar to the tensile strength of rock (less than ~10 MPa ref.[8]). Additionally, considering that our calculations of excess pressure are conservative, it is likely that tensile failure and dike propagation in warm solid iron is possible. Another possibility is that deeper pockets of Fe-FeS melt could migrate to the core mantle boundary as diapirs[7]. Such a possibility could make ferrovolcanism a viable process even if core growth proceeds concentrically inward rather than by growth of dendrites.

The mantle penetration height, $t$, is calculated by setting the excess pressure (e.g. Figure 1) equal to the negative buoyancy force of sulfur enriched core material being intruded into the mantle $P_e = hg(\rho_{Fe} - \rho_{FeS}) = tg\,(\rho_{FeS} - \rho_{mantle})$, where the density of the mantle $\rho_{mantle} = 3300$ kg/m³. Thus, $t = h(\rho_{Fe} - \rho_{FeS})/(\rho_{FeS} - \rho_{mantle})$, and $t$ is independent of $g$ and core size. This method is the same approach used by refs.[7,30]. If the interconnected melt was located deeper in the core such that the top of the melt was at a depth D from the core mantle boundary the mantle penetration would instead be given by $t = (h + D)(\rho_{Fe} - \rho_{FeS})/(\rho_{FeS} - \rho_{mantle})$, which is always larger than the penetration height shown in figure 2. These calculations ignore the small effect of increasing gravitational acceleration with distance from the center of the planetesimal. For reference, the gravitational acceleration at the surface of a 100-km-radius core is 0.196 m/s² while the gravitational acceleration at the surface of 200-km-radius planetesimal with a 100-km-radius core is 0.211 m/s². Thus, the height-averaged gravitational acceleration experienced by a melt propagating 100 km to the surface of such a 200-km-radius planetesimal would be about 4% higher than our simple calculation suggests. We argue the benefit of presenting results in Figure 2 that are independent of $g$ and core size outweighs the benefit of slightly more accurate calculations that include this minor effect. We note that for a given excess pressure, the mantle penetration height depends on gravity. This is important when considering terrestrial dikes that are efficiently arrested when reaching a low-density layer[31]. If a terrestrial dike were able to propagate 1 km into low density material, for the same density contrast and excess pressure, a lunar dike could propagate 6 km and a dike originating from a 100-km radius core could propagate 50 km.

As previously noted penetration height is sensitive to pre-existing stress gradients. Generally, a cooling and contracting planetesimal is expected to be in a state of compression, which can hinder dike propagation[32]. Intermittent thrust faulting can relieve these compressive stresses[32] and possibly increase excess pressures aiding the ascent of ferromagma. Additionally, locally favorable stress states (due to, for example, topography) may aid a modest amount of extrusion even when a cooling and contracting planet is not experiencing widespread volcanism as a whole because it is in a globally compressive stress state. This process has been inferred to occur on Mercury[33].

There is some indication that pallasite cooling rates are higher for melts that are less evolved and contain less iridium[12]. Because a more evolved Fe-FeS melt is more enriched in sulfur, we might expect more evolved melts to more effectively penetrate the mantle, come closer to the surface of the planetesimal, and have higher cooling rates. However, this idea ignores several factors that can influence the ascent of ferromagmas, which require more detailed consideration of the thermal and physical evolution of planetesimals. One important effect to consider is how the vertical height of interconnected melt affects the excess pressures and how lateral variation in mantle thickness could affect cooling rate. It is possible that locations where the mantle is thinner will result in a larger $h$ resulting in intrusion of less evolved ferromagma to height similar to that of a more evolved ferromagma. In such a region of locally thin mantle, the cooling rate would be higher for the same mantle penetration $t$. Another possibility is that the vertical height of interconnected melt decreases as the core solidifies, as thermal models of ref.4 indicate, which could cause later more evolved ferromagmas to have lower penetration heights $t$. It is also likely that as planetesimals cool and contract ascent becomes more difficult even though magmas may be more evolved and have higher sulfur contents. Finally, volatiles and light elements play an important role in magmatic processes, including lowering the bulk density of magma. It is possible that primordial volatiles other than sulfur, such as nitrogen and carbon monoxide, could aid ascent of early, less evolved ferromagmas in this way[11], but become depleted for more evolved ferromagmas.

We calculate the density of Psyche under the assumption that it is a triaxial ellipsoid with radii 274, 231, and 176 km. These values come from estimates of Psyche's shape derived from ground-based observations[5]. We consider a two-layer model, with an inner metallic core of density 7500 kg/m$^3$ and an outer silicate layer (which we call the mantle) with densities ranging from 2500–3500 kg/m³. The mantle has a uniform thickness, which is a free parameter ranging from 0–80 km. This two-layer structure with uniform thickness is simple, and Psyche is almost certainly more complex as a result of factors like impact-induced mixing and topography, which has been putatively detected[6]. Nonetheless, this model serves to estimate how much rocky material would be needed to match density estimates[5].

**Data Availability:**

The data that support the plots within this paper and other findings of this study are available from the corresponding author upon reasonable request.

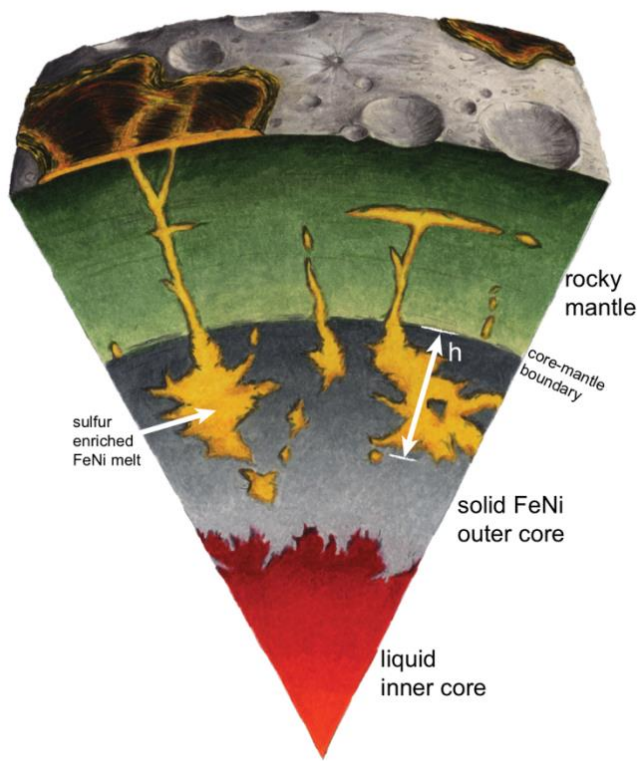

**Figure 1: Structure of a differentiated planetesimal that may experience ferrovolcanic eruptions of sulfur-rich FeNi melt.** The vertical extent $h$ of an interconnected sulfur-enriched FeNi melt within a solid outer core is illustrated. Core size, mantle thickness, and vertical extent of interconnected sulfur-enriched FeNi melt are treated as independent variables in our calculations. Illustration by James Tuttle Keane (Caltech).

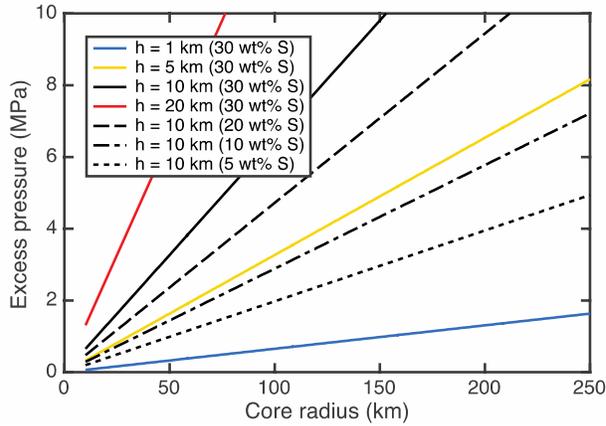

**Figure 2: Excess pressure associated with the buoyancy of sulfur-enriched iron melts surrounded by dense solid iron as function of core radius**. The colored lines represent different vertical extents, $h$, of an interconnected region of melt. The black dotted and dashed lines show the effect of sulfur enrichment at a fixed value of h=10 km. The eutectic composition of a sulfur enriched iron melt is 31 wt% sulfur (weight percent). Core radius is total core radius including both solid and liquid portions.

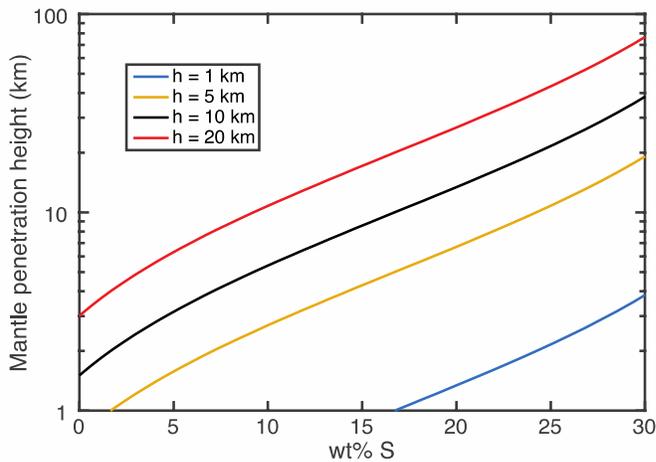

**Figure 3: Mantle penetration height of ferromagmatic dikes as a function of sulfur content.** The colored curves represent different vertical extents, $h$, of an interconnected region of partial melt. Note that mantle penetration does not depend on core size or gravitational acceleration (see Methods).

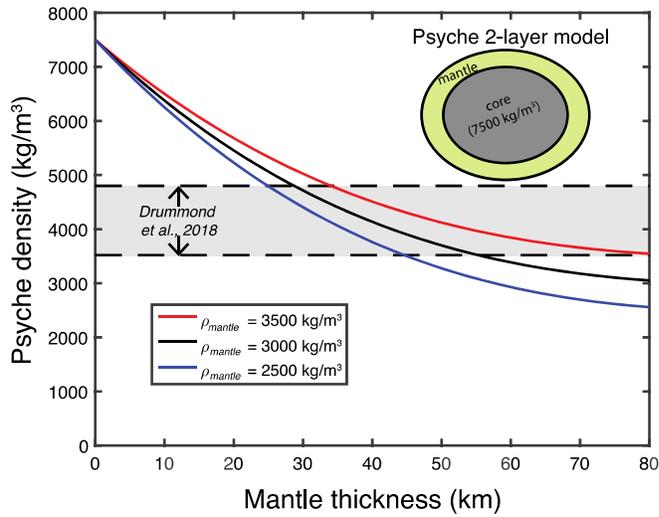

**Figure 4: Modeled density of Psyche assuming a two-layer structure of a metal core and a rocky mantle compared with the observed density.** Colored curves assume different densities for the Psyche's mantle ($\rho_{mantle}$) as indicated. The gray box between the dashed lines is the range of densities that are consistent with observations[5].